\documentclass[twoside]{reporteq}
\usepackage{svcon2e}
\usepackage{makeidx}
\usepackage{amssymb}
\usepackage{algorithmic}
\usepackage{latexsym}
\usepackage{amsmath}
\usepackage{graphicx}
\usepackage{amsthm}

\newcommand{\ed}{\end{document}}
\makeindex

\newcommand{\BC}{\mathbb{C}}
\newcommand{\BR}{\mathbb{R}}

\newtheorem{theorem}{Theorem}

\newcommand{\A}{{\cal A}}

\newcommand{\beq}{\begin{equation}}
\newcommand{\eeq}{\end{equation}}
\newcommand{\bp}{{\bf p}}
\newcommand{\bq}{{\bf q}}
\newcommand{\bx}{{\bf x}}

\newcommand{\bn}{{\bf n}}

\begin{document}

\chapter{Quantum Fractals. Geometric\break
modeling of quantum jumps\break with conformal maps\ }
\chapterauthors{Arkadiusz Jadczyk}
{ {\renewcommand{\thefootnote}{\fnsymbol{footnote}}
\footnotetext{\kern-15.3pt AMS Subject Classification: 15A66, 28A80,
81P99 .} }}

\begin{abstract}
Positive matrices in $SL(2,\BC)$ have a double physical
interpretation; they can be either considered as ``fuzzy
projections" of a spin $1/2$ quantum system, or as Lorentz boosts.
In the present paper, concentrating on this second interpretation,
we follow the clues given by Pertti Lounesto and, using the
classical Clifford algebraic methods, interpret them as conformal
maps of the ``heavenly sphere" $S^2.$ The fuzziness parameter of the
first interpretation becomes the ``boost velocity" in the second
one. We discuss simple iterative function systems of such maps, and
show that they lead to self--similar fractal patterns on $S^2.$ The
final section of this paper is devoted to an informal discussion of
the relations between these concepts and the problems in the
foundations of quantum theory, where the interplay between different
kinds of algebras and maps may enable us to describe not only the
continuous evolution of wave functions, but also quantum jumps and
``events" that accompany these jumps.\footnote{Paper dedicated to
the memory of Pertti Lounesto}
\\
\noindent {\bf Keywords: } Clifford algebras, conformal maps,
iterated function systems, quantum jumps, quantum fractals.\par
\end{abstract}

\pagestyle{myheadings} \markboth{Arkadiusz Jadczyk}{Quantum
Fractals}
\section{Introduction}
\begin{figure}[!htb]
\centerline{\includegraphics[width=10cm]{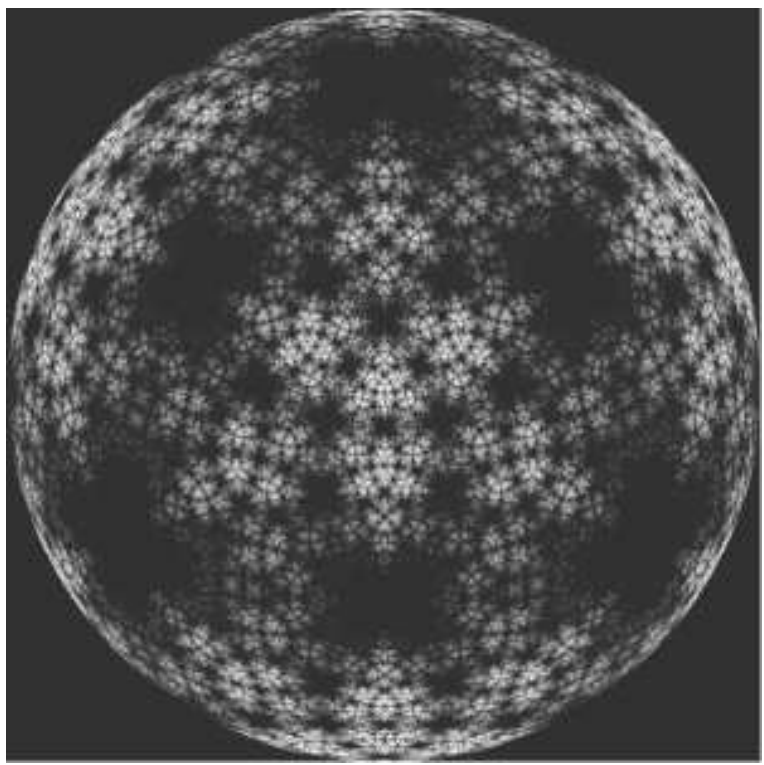}}
\caption{Quantum fractal based on eight vertices of a cube inscribed
in the unit sphere $S^2$. 100,000,000 points obtained by a random
choice of the initial point, followed by the application of randomly
chosen conformal maps, with place--dependent probabilities $p_i$
given by the formula (\ref{eq:pi}), from among eight maps defined by
unit vectors $\bn_i$--s situated at the eight vertices of a cube.
View from above one of the vertices. Other closest three vertices
are located at 60, 180 and 240 degrees. The dark areas are those
that are (almost) never visited. The white areas are those that are
frequently visited. The pattern shows distinct self--similarity -
circles with circles. The details of algorithm are described in in
Sec. \ref{sec:qf} } \label{fig:Cube}
\end{figure}

Let ${\bf B}^3=\{{\bf q}\in \BR^3:\, {\bf q}^2\leq 1\}$ be the unit
ball in $\BR^3$ and let $S^2=\{{\bf p}\in \BR^3:\, {\bf p}^2=1\}$ be
the unit 2--sphere, that is the boundary of ${\bf B}^3.$ Every
$\bq\in \BR^3$ determines a map $\phi_{\bq}: S^2\rightarrow S^2$
through the formula: \beq \phi_\bq(\bp)\doteq
\frac{(1-\bq^2)\bp+2(1+\bq\cdot\bp)\bq}{1+\bq^2+2\bq\cdot\bp}.\label{eq:1}\eeq
The formula (\ref{eq:1}) came naturally when discussing quantum
jumps of a state of a spin $\frac{1}{2}$ particle \cite{ja94a}.
\footnote{Notice that the formula makes also sense if $\bq^2>1,$ but
in this case the $\phi_{\bq}$ is equivalent to the map
$\phi_{\bq/{\bq^2}}$ followed by the inversion in the plane
perpendicular to $\bq.$} During the 6-th ICCA Conference,  Pertti
Lounesto \cite{lounesto1} conjectured that the maps $\phi_\bq,\,
\bq\in {\bf B}^3,$ are conformal maps in that they preserve angles
between vectors tangent to the sphere $S^2,$ and he checked it
numerically on randomly chosen tangent vectors using CLICAL
\cite{lounesto2}. Interesting patterns arise when the transformation
$\phi_{\bq}$ is iterated, that is applied many times, using
different, symmetrically distributed $\bq$'s. For instance, taking
eight vectors $\bq_i,\, i=1,2,\ldots ,,8,$ pointing from the origin
to the eight corners of a cube inscribed in the unit sphere, all
$\bq_i$'s of length, say, $\Vert \bq_i\Vert =0.74,$ we get the
pattern shown in Fig.\ \ref{fig:Cube}.
\subsection{Iterated maps. Hausdorff distance, contractions, and attractor set.}
Let $(X,d)$ be a complete metric space.  In our examples $X$ will
be a compact subset of the real plane $\BR^2$ or a
$2$--dimensional sphere $S^2=\{(x,y)\in\BR^2:\, x^2+y^2=1\},$
which is also a complete metric space when endowed with the
geodesic distance function $d(x,y)$ being the arc length along the
great circle connecting $x$ and $y.$ Let ${\cal H}(X)$ be the set
of all non--empty compact subsets of $X.$ A distance $h(Y,Z)$
(Hausdorff metric) between any two sets $Y,Z\subset X$ can be
defined as follows. First define the distance between any point
$x\in$ and any $Y\in {\cal H}(X)$ by
$$d(x,Y)=\{ \min d(x,y):\, y\in Y\}.$$ Then, for any $X,Y\in{\cal
H}(X)$ define the distance $d(Y,Z)$ {\em from}\ set $Y$ {\em to}\
set $Z$ by the formula
$$d(Y,Z)=\max\{ d(y,Z):\ y\in Y\}.$$ The formula for $d(Y,Z)$
is not symmetric in $Y$ and $Z.$ Therefore one defines the
Hausdorff distance $h(Y,Z)$ as the $\max$ of the two:
$$h(Y,Z)=\max (d(Y,Z),d(Z,Y)).$$ It can be shown that $h(Y,Z)$ is a metric
on ${\cal H}(X).$ The definition of the Hausdorff distance is not
very intuitive. There is an intuitive way to understand it: two
sets are within Hausdorff distance $r$ from each other if and only
if any point of one set is within distance $r$ from some point of
the other set. From the fact that $X$ is also a complete metric
space it can be then shown that ${\cal H}(X)$ endowed with the
Hausdorff metric is a complete metric space, and therefore every
Cauchy sequence $Y_n\in {\cal H}(X)$ has a limit in ${\cal H}(X).$
This property is crucial in proving the existence of attractor
sets in studies of iterated function systems. A map
$f:X\rightarrow X$ is a {\em contraction}\, if there exists a
constant $s,$ $0<s<1,$ called the {\em contraction factor}, such
that $d(f(x),f(x^\prime))<s\cdot d(x,x^\prime )$ for any two
different points $x,x^\prime \in X.$ The so called Contraction Map
Theorem states that in a complete metric space every contraction
map $f$ has a unique fixed point $x_0,$ i.e. such that
$f(x_0)=x_0.$ Moreover, for any initial point $x\in X$ the
sequence $x_n=f^{(n)}(x),$ where $f^{(n)}=f\circ f\circ \ldots
\circ f$ ($n$ times), converges to $x_0. $ Let now $f_1,f_2,\ldots
f_n$ be contraction maps $f_k: X\rightarrow X,$ $k=1,2,\ldots n,$
with contraction factors $s_k.$. Then we can define a map $F$
acting on subsets $Y\subset X$ by the formula:
$$F(Y)= f_1(Y)\cup f_2(Y)\cup\ldots \cup f_n(Y)$$ where $Y\in
{\cal H}(X)$ and $f_k(Y)$ is the image of the set $Y$ under the map
$f.$\footnote{$F$ is called the {\em Hutchinson operator.}} It can
be shown that $F$ restricts to a map $F: {\cal H}(X)\rightarrow
{\cal H}(X)$, and that this map is a contraction with the
contraction factor $s=\max(s_1,\ldots ,s_k).$ It follows from the
Contraction Mapping Theorem that $F$ has a unique fixed point, in
that there is a unique compact subset $Y_0\subset X$ with the
property that
$$Y_0=\bigcup_{k=1}^n f_k(Y_0).$$ This set $Y_0$ is called an {\em
attractor set }\ for the {\em Iterated Function System}\
consisting of the family $(f_1,\ldots ,f_n).$ Finding a numerical
approximation to the attractor set needs lot of computation. Even
when we start with a one--point set, its image under $F^{(k+1)}$
may have $n^k$ points. In cases like that moving to probabilistic
algorithms may drastically reduce the need for computing
resources. Quantum theory, that is probabilistic in nature, offers
naturally examples of Iterated Function Systems with probabilities
assigned to the maps $f_i.$ Such a system is called ``IFS with
probabilities" \cite[Ch. 9.1]{barnsley}. The simplest example is
provided by three affine maps with Sierpinski triangle as the
attractor set.
\subsection{The Sierpinski triangle.}
\begin{figure}[h!]
  \centering
  \scalebox{0.25}{\includegraphics{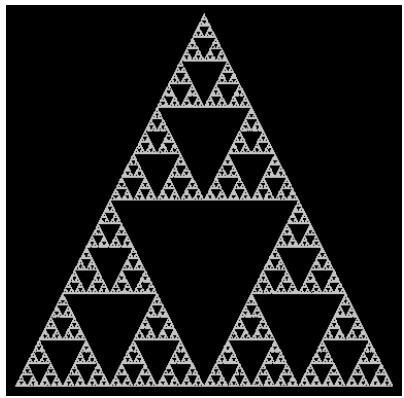}}
  \caption{Sierpinski triangle. The attractor set of three non--commuting affine contractions. }\label{fig:sierpinski}
\end{figure}
An affine transformation of $\BR^2$ is of the form $x\mapsto
Ax+b,$ where $A$ is a $2\times 2$ matrix and $x,b\in\BR^2.$ It is
often convenient to represent such a transformation as a $3\times
3$ matrix $${\tilde A}=\begin{pmatrix}A& a\cr
0&1\cr\end{pmatrix}$$ acting on $\BR^2$ embedded in $\BR^3$ as
follows:
$$\begin{pmatrix}x\cr 1\cr\end{pmatrix}\mapsto
\begin{pmatrix}A& a\cr0&1\cr\end{pmatrix}\begin{pmatrix}x\cr
1\cr\end{pmatrix}=\begin{pmatrix}Ax+a\cr 1\cr\end{pmatrix}.$$ An
affine transformation ${\tilde A}$ is a {\em contraction}\, if for
each $0\neq x\in \BR^2$ we have that $\Vert Ax\Vert < \Vert
x\Vert.$ Consider now three affine transformations ${\tilde
A}[i],\quad i=1,2,3$ defined by $$ {\tilde
A}[i]=\begin{pmatrix}0.5& 0 &x[i]\cr 0& 0.5& y[i]\cr0& 0&
1\cr\end{pmatrix}$$ where $x[1]=y[1]=0,$ $x[2]=0.5,\, y[2]=0,$
$x[3]=0.25,\, y[3]=0.5.$ The transformations ${\tilde A}[i]$ do
not commute. For instance ${\tilde A}[2]{\tilde A}[1]-{\tilde
A}[1]{\tilde A}[2]$ is a translation by $0.25$ in the $x$
direction. They are also contractions, and they map the square
$X=\{(x,y):\, 0 \leq x\leq 1,\, 0\leq y \leq 1 \},$ into itself
(cf. \cite{barnsley}[Ch. 3.7]). The probabilistic algorithm goes
as follows: one starts with an arbitrary initial point $x_0$ and
applies to it one of the three transformations $f_i,$ selected
randomly, each with the probability $p_i=1/3.$ One gets a new
point $x_1.$ Then one of the transformations, again selected
randomly, is applied to $x_1$ to produce $x_2,$ etc. Each point is
being plotted. The result of 100,000 transformations is presented
in Fig.\ \ref{fig:sierpinski}. \section{M\"obius transformations
of $S^2.$}
\subsection{Notation\label{sec:notation}}We denote
by $E_{(r,s)}$ the real vector space $\BR^n,$ $n=r+s,$ endowed
with the quadratic form $q(x)$ of signature $(r,s).$
$E_n=E_{(n,0)}$ is the standard $n$--dimensional Euclidean space.
The Clifford algebra of $E_{(r,s)}$ is denoted by $C(E_{(r,s)}),$
and the Clifford map $E_{(r,s)}\ni x \mapsto \phi(x)\in
C(E_{(r,s)})$ satisfies $\phi(x)^2=q(x)I.$ $x$ and $\phi(x)$ are
often identified. The principal automorphism of $C(E_{(r,s)})$ is
denoted by $\pi$ and is determined by $\pi(x)=-(x),\, x\in
E_{(r,s)},$ while the principal anti--automorphism $\tau$ is
determined by $\tau(x)=x.$ Their composition $\nu$ is also denoted
as $\nu(a)={\tilde a}$ and is the unique anti--automorphism
satisfying ${\tilde x}=-x$ for all $x\in E_{(r,s)}.$ $\BC(n)$
(resp $\BR(n)$) will denote the
algebra of complex (resp. real) matrices $n\times n.$\\
The Pauli spin matrices
$\sigma_0,\,\sigma_1,\,\sigma_2,\,\sigma_3$ are given by
$$\sigma_0\:=\:\left(\begin{array}{cc}1&0\\0&1\end{array}\right)\quad\sigma_1\:=\:\left(\begin{array}{cc}
0& 1\\1& 0 \end{array}\right)\quad
\sigma_2\:=\:\left(\begin{array}{cc} 0& -i\\i& 0
\end{array}\right)\quad \sigma_3\:=\:\left(\begin{array}{cc} 1&
0\\0& -1 \end{array}\right).$$ We have
$$X=\begin{pmatrix}x^0+x^3& x^1-ix^2\cr x^1+ix^2& x^0-x^3\cr\end{pmatrix},$$ thus $\det (X)=(x^0)^2-{\bf x}^2,$
and therefore
an isomorphism of the Minkowski space $E_{(1,3)}$ with the
$2\times 2$ hermitian matrices $X=X^\star.$ The inverse map
$X\mapsto (x^\mu)$ is given be
$x^\mu=\frac{1}{2}\mbox(Tr)(\sigma_mu X).$ It is easy to verify
that the Pauli matrices satisfy the following relations:
$$\sigma_\mu^\star=\sigma_\mu,\qquad $$
$$\sigma_k\sigma_l=i\epsilon_{klm}\sigma_m,$$
$$\frac{1}{2}\mbox{Tr}(\sigma_\mu\sigma_\nu)=\delta_{\mu\nu},$$
$$\frac{1}{2}\mbox{Tr}(\sigma_k\sigma_l\sigma_m)=i\epsilon_{klm},$$
$$\frac{1}{2}\mbox{Tr}(\sigma_j\sigma_k\sigma_l\sigma_m)=\delta_{jk}\delta_{lm}+\delta_{jm}\delta_{kl}-\delta_{jl}\delta_{km},$$
where $\mu ,\nu =0,1,2,3,$ and $j,k,l,m =1,2,3.$ The map $E_3\ni
{\bf x}\mapsto \sigma({\bf
x})=x^1\sigma_1+x^2\sigma_2+x^3\sigma_3$ is a Clifford map from
$E_3$ to $\BC(2),$ and $\BC(2),$ as a {\em real} algebra, can be
considered as the Clifford algebra of $E_3.$

\subsection{$SL(2,\BC)$ as the group of M\"obius transformations
of $S^2.$} We will be interested in the particular case of $n=2,$ in
which case the connected component of identity of the conformal
group $Conf(\BR^2)$ is isomorphic to the ortochronous Lorentz group
$SO_{+}(3,1).$ If we identify $S^2$ with the compactified complex
plane $\BC\cup{\infty},$ then conformal transformations form
$Conf_+(\BR^2)$ can be conveniently realized by complex homographies
$\BC \ni z\mapsto\frac{az+b}{cz+d}$ (\cite{angles}[Exercise 2.13.1].
For our purposes it will be more convenient to use the group
$Spin(1,3)$ realized as $Sp(2,\BC)\approx SL(2,\BC).$ We will start
with describing the isomorphism of $Spin(1,3)$ to $SL(2,\BC)$
following the simple method given by Deheuvels in
\cite{deheuvels}[Ch. X.6]

 Every Hermitian $2\times 2$ matrix $X$ can be uniquely
represented as $$X=x^\mu\sigma_\mu$$, with $x^\mu$ real, and where
$\sigma_\mu$ are the Pauli matrices. For every $2\times 2$ matrix
$A$ define $A^{\checkmark}\doteq CA^tC^{-1},$ where $A^{t}$ is the
transposed matrix and
$$C=\begin{pmatrix}0&-1\cr 1&0\cr\end{pmatrix}.$$ Then $A\mapsto
A^\checkmark$ is an anti--involution of the algebra $\BC(2)$ and
we have
$$A^\checkmark A= AA^\checkmark = \det(A)I$$ for all $A\in\BC(2).$
In particular, $A\in SL(2,\BC)$ if and only if $A^\checkmark
=A^{-1}.$ Notice that the anti--automorphisms $A\mapsto A^\star$
and $A\mapsto\A^\checkmark$ commute. Their composition denoted by
$A\mapsto {\tilde A}=C{\bar A}C^{-1}$ is an involutive
automorphism of the {\em real} algebra $\BC(2),$ and it coincides
with the automorphism $A\mapsto{\tilde A}$ if $\BC(2)$ is
considered as the Clifford algebra of $E_3$ with the Clifford map
${\bf x}\mapsto \sigma({\bf x}).$ Notice that for $A\in SL(2,\BC)$
we have ${\tilde A}=A^\star.$ It follows that the map $x\mapsto
\phi(x)$ defined by
$$\phi(x)=\begin{pmatrix}0&X\cr X^\checkmark&0\end{pmatrix}$$ is a
Clifford map from $E_{1,3}$ into the algebra $\BC(4)$ od complex
$4\times 4$ matrices. It is shown in \cite{deheuvels}[Th\'eor\`eme
X.6] that $SL(2,\BC)$ can be identified then with the group
$Spin(1,3)\subset\BC(4)$ via the mapping $$SL(2,\BC)\ni
A\mapsto\begin{pmatrix}A&0\cr 0&\tilde A\cr\end{pmatrix}.$$ The
action $Spin(1,3)$ on $E_{(1,3)}$ can be then easily computed in
terms of $SL(2,\BC)$ matrices:
$$\begin{pmatrix} A&0\cr 0&{\tilde A}\end{pmatrix}\begin{pmatrix}0&
X\cr X^\checkmark &0\end{pmatrix}\begin{pmatrix}A^{-1}&0\cr 0&
{\tilde A}^{-1}\end{pmatrix}=\begin{pmatrix}0&X^\prime\cr
{X^\prime}^\checkmark &0\end{pmatrix},$$ where $X^\prime=AX{\tilde
A}^{-1}=AXA^\star.$ If $X=x^\mu\sigma_\mu$ then the map is
accomplished by a Lorentz matrix ${\Lambda(A)^\mu}_\nu$ via
$$x^{\prime\mu}={\Lambda(A)^\mu}_\nu\, x^\nu.$$
\vskip10pt \noindent{\bf Note:} It is sometimes convenient to
parametrize $GL(2,\BC)$ by complex Minkowski space coordinates
$a^\mu\in\BC,$ via $ A=a^\mu\sigma_\mu.$ It easily follows that
$A\in SL(2,\BC)$ if and only if $a^2=(a^0)^2-{\bf a}^2=1.$ Using
the formulas of section \ref{sec:notation} we can express the
components of the Lorentz matrix ${\Lambda(A)^\mu}_\nu$ through
the complex coordinates $a^\mu$ of $A$ as follows: $$
{\Lambda^0}_0=|a^0|^2+|{\bf a}|^2,$$
$${\Lambda^0}_j=2\Re({\bar a}^0a^j)+i\epsilon_{jkl}a^k{\bar
a}^l={\Lambda^j}_0,$$
$${\Lambda^j}_k=(a\cdot{\bar a})\, \delta^j_k+2\Re (a^j{\bar a}^k)+2\Im
({\bar a}^0a^l)\, \epsilon_{jkl}.$$ \vskip10pt  In order to
describe explicitly the action of $SL(2,\BC)$ on $S^2$ it is
convenient to embed $S^2$ in $E_{(1,3)}$ via $x^0=1$ section of
the light--cone $x^2=0.$ That is we identify $S^2$ with the
boundary of the unit ball $S^2=\{{\bf x}\in \BR^3:{\bf
x}^2=1\}=\{x=(x^0,{\bf x})\in E_{(1,3)}: x^0=1, x^2=0\}.$ Given a
unit vector $\bx\in S^2\subset \BR^3,$ we associate with it the
null vector $x=(1,\bx)\in E_{(1,3)},$ and therefore the matrix
$$X=\sigma_0+x^i\sigma_i=\begin{pmatrix}1+x^3&x^3-ix^2\\x^3+ix^2&1-x^3\end{pmatrix}.$$
The matrix $X$ is positive and of determinant zero. The $SL(2,\BC
)$ transformed matrix \beq X^\prime = AXA^\star\label{eq:axa}\eeq
is also positive and of determinant zero. Therefore it represents
another future oriented, null vector $x^\prime ,$ that corresponds
to a unique vector $\bx^\prime\in S^2$. In our application we will
be interested in special conformal transformations of $S^2,$
namely those generated by ``pure boosts'' of $SL(2,\BC ).$ By the
polar decomposition theorem every matrix $A\in SL(2,\BC )$ can be
uniquely decomposed into a product of a unitary and a positive
matrix - both of determinant one. Unitary matrices represent
three--dimensional rotations, while positive matrices represent
special Lorentz transformations (boosts).\footnote{It is important
to notice that the isomorphism of $Spin(1,3)$ and $SL(2,\BC )$ is
not a natural one. It depends on a chosen Lorentz frame. Therefore
the splitting of a group element into the product of a pure
rotation and a boost also depends on the chosen Lorentz frame. }
The most general form of a positive $SL(2,\BC )$ matrix is \beq
P(\bn,\alpha)=c(I+\alpha\sigma(\bn )), \eeq where $\bn\in S^2$ is
a unit vector (the boost direction), and $0<\alpha=v/c<1$ is the
``boost velocity". \footnote{The constant $c$ should be chosen to
be $c=(1/\sqrt(1-\alpha^2)),$ to assure that the determinant is
one, but we will put $c=1,$ because the constant factor cancels
out anyway when going to the induced action on $S^2.$} Sometimes
we will simply write $P(\bq),$ instead of $P(\bn,\alpha),$ putting
$\bq=\alpha\bn :$ \beq P(\bq )=(I+\sigma(\bq )).\eeq  In the limit
of $\alpha=1,$ which corresponds to ``the velocity of light'' $P$
degenerates into a projection operator, and we have $P(\bx)=X,$
where $X$ represents the null vector $x=\{x^\mu\}=(1,\bx),$
$\bx\in S^2.$ Since $P(\bq)=P(\bq )^\star,$ the action of the
boosts $P(\bq )$ on vectors $\bx\in S^2$ given by the Eq.
(\ref{eq:axa}) can be found from the formula:

\beq P(\bq )P(\bx)P(\bq )=\lambda (\bq,\bx)
P(\bx^\prime).\label{eq:ppp}\eeq A straightforward calculation
gives \beq \lambda(\bq,\bx)=\frac{1+\bq^2+2\,\bq\cdot\bx
}{4},\label{eq:l}\eeq \beq
\bx^\prime=\frac{(1-\bq^2)\bx+2(1+\bq\cdot\bx
)\bx}{1+\bq^2+2\,\bq\cdot\bx }.\label{eq:xprime}\eeq Therefore we
recover the formula (\ref{eq:1}) as coming from the special
conformal transformation in the group $Spin(1,3).$ The crucial
point in the above is to notice that $S^2$ is the one--point
compactification of $E_2$ (the Riemann sphere), and that
$E_{(3,1)}=E_{(2+1,0+1)},$ so that $Spin(1,3)=Spin(3,1)$ is the
covering group of the conformal group for $E_2$ and $S^2.$
\subsection{The geometrical meaning of the coefficient $\lambda(\bq,\bx).$}
The numerical coefficient $\lambda(\bq,\bx)$ in the formula
(\ref{eq:ppp}) is not important for the transformation
$\bx\mapsto\bx^\prime.$ Yet in the studies of iterated function
systems not only the transformations themselves, but also the
probabilities assigned to the transformations play an important
role. For instance in Ref.\cite [Chapter 6.3, p. 329]{peitgen} we
find that for affine contractions it is advisable to choose the
probabilities of maps to be proportional to the determinants of
their linear parts. In our case the maps are M\"obius
transformations of $S^2,$ and they are not contractions. In fact
these maps contract some regions while expanding other regions. Is
there a ``natural'' choice of probabilities, and can we use the
place dependent factors $\lambda(\bq,\bx)$ for determining the
natural choice of probabilities? The answer is ``yes'', though the
exact formula is not at all evident. In \cite{jad06c} it is shown
that by choosing $\lambda(\bq,\bx)$' as the relative probabilities
of M\"obius transformations (\ref{eq:xprime}), the iterated
function system leads to a Markov semigroup that is {\em linear}.
Moreover, denoting by $dS$ the rotation invariant area element of
$S^2,$ we find that this area changes as the result of the
M\"obius transformation (\ref{eq:xprime}) according to the
formula: \beq \frac{dS^\prime}{dS}(\bx
)=\frac{(1-\bq^2)^2}{(1+\bq^2+2\bq\cdot\bx)^2}.\eeq To visualize
the mapping, let us assume that $\bq=\alpha \bn,$ and that the
vector $\bn$ is along the $z$ axis: $\bn=(0,0,1).$ Then all the
region of the sphere above the critical value of $z=-\alpha$ is
contracted into the region of the sphere above $z=\alpha,$ and the
region of the sphere below $z=-\alpha$ is expanded into the region
of the sphere below $z=\alpha.$ The relative probability
$\lambda(\bq,\bx)$ of choosing the M\"obius map determined by
$\bq=\alpha \bn$ is highest, $\lambda_{max}=(1+\alpha)^2,$ for
$\bx$ parallel to $\bn$ and has the minimum,
$\lambda_{min}=(1-\alpha)^2$ for $\bx$ antiparallel to $\bn.$ At
the critical value of $z=-\alpha,$ we have $\lambda=1-\alpha^2,$
which is the geometrical mean of $\lambda_{max}$ and of
$\lambda_{min}.$
\section{Quantum Fractals\label{sec:qf}}
In order to implement an IFS with M\"obius maps of the type that
we have discussed, we need $N$ unit vectors $\bn_i,\,
i=1,\ldots,N,$ and $N$ constants $\alpha_i,$ $0<\alpha_i<1.$ Each
vector $\bn_i$ determines the direction, while each constant
$\alpha_i$ determines the velocity of the Lorentz boost that
implements the M\"obius transformation $\phi_{\bq_i}$ of $S^2:$
\beq \phi_{\bq_i}(\bp)\doteq
\frac{(1-\bq^2)\bp+2(1+\bq_i\cdot\bp)\bq_i}{1+\bq^2+2\bq_i\cdot\bp},\eeq
with $\bq_i=\alpha_i\bn_i.$. The probability $p_i(\bx)$ of
selecting the map $\phi_i=\phi_{\bq_i}$ is then given by: \beq
p_i(\bx)=\frac{\lambda(\bq_i,\bx)}{\sum_{j=1}^N
\lambda(\bq_j,\bx)}.\eeq Inspecting the formula (\ref{eq:l}) we
see that the denominator $\sum_{j=1}^N \lambda(\bq_j,\bx)$
simplifies essentially if all $\alpha_i$ are the same:
$\alpha_i=\alpha,\, i=1,\ldots ,N,$ and the vectors $\bn_i$
average to zero: $\sum_{i=1}^N \bn_i = 0.$ In this case the
formula for probabilities $p_i(\bx)$ simplifies to: \beq
p_i(\bx)=\frac{1+\alpha^2+2\alpha\,
\bn_i\cdot\bx}{N(1+\alpha^2)}.\label{eq:pi}\eeq
\subsection{Pseudocode for generation of M\"obius IFS }
\begin{figure}[hbtp]
  \centerline{\includegraphics[width=11cm]{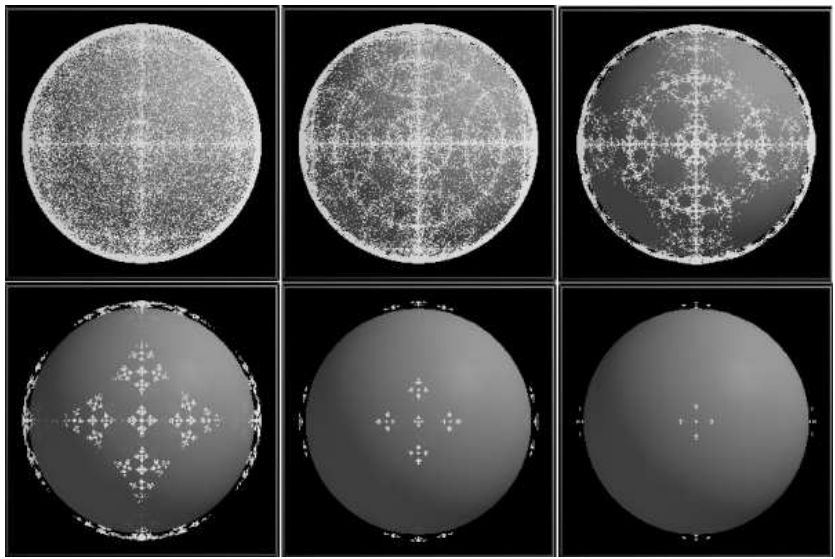}}
  \caption{Quantum Fractal created with six vertices forming a regular octahedron, for $\alpha=0.4,0.5,\ldots ,0.9.$ }
  \label{fig:octa}
\end{figure}

In order to implement the IFS described above we first need to
choose a set of unit vectors $\bn_i,$ and a value of the constant
$\alpha.$ For instance, to create the picture, like that in Fig.\
\ref{fig:Cube}, we have chosen $\alpha=0.71,$ and the vectors
$\bn_i$ as pointing to the eight vectors of the cube inscribed
into the unit sphere, with
one of the vertices at the north pole:\\ $\bn_1= (0, 0, 1),$
$\bn_2=(2\sqrt{2}/3,0,1/3),$ $ \bn_3=(-2\sqrt{2/3}, 0, 1/3),$\\
 $\bn_4=(-\sqrt{2}/3, -\sqrt{2/3}, 1/3),$ $\bn_5=(\sqrt{2}/3,
\sqrt{2/3}, -1/3),$\\ $\bn_6=(\sqrt{2}/3, -\sqrt{2/3}, -1/3),$
$\bn_7=(-2\sqrt{2}/3, 0, -1/3),$ $\bn_8=(0, 0, -1).$
 The following pseudocode
describes now the generation of an IFS with M\"obius
transformations: \vskip10pt
\begin{algorithmic} \STATE (select initial $\bx$) \STATE
$\bx\leftarrow\bx_0$\STATE (choose $imax,$ for instance) \STATE
$imax\leftarrow 10000000$ \STATE $icount \leftarrow 0$
\WHILE{$icount<imax$} \STATE $icount\leftarrow icount+1$ \STATE
(select one of the maps $\Phi_i$) \STATE (first initialize
probability) \STATE $p\leftarrow 0$ \STATE (initialize maps
counter) \STATE $i\leftarrow 0$ \STATE (choose a random number
$0<r<1$) \STATE $r\leftarrow \mbox{random}(1)$ \REPEAT \STATE
$i\leftarrow i+1$ \STATE $p\leftarrow p+p_i(\bx)$ \UNTIL{$p>r$}
\STATE (the map $\phi_i$ is now selected, apply it) \STATE
$\bx\leftarrow\phi_i(\bx)$\ENDWHILE
\end{algorithmic}
\vskip10pt To create a graphic representation, such as in Fig.\
\ref{fig:Cube}, we project the upper hemisphere onto the plane
$(x,y),$ and divide the unit square of this plane into $r_x\times
r_y$, for instance $600\times 600,$ rectangular cells, each cell
being represented by one pixel on the screen. We associate a
counter $c[ix][iy]$ with each of the cells $(ix,iy)$, initialize
all counters to $0,$ and count points $\bx=(x,y,z)$ that fall into
the cell: \vskip10pt
\begin{algorithmic}
\STATE $ delta_x\leftarrow 2.0/r_x;\, delta_y\leftarrow 2.0/r_y$
\STATE $ix\leftarrow \mbox{round}((x-(-1.0))/delta_x);\,
iy\leftarrow \mbox{round}((y-(-1.0))/delta_y)$ \STATE (increase
counter $c[i][j]$ by one:) \STATE $c[ix][iy]\leftarrow
c[ix][iy]+1$
\end{algorithmic}
\vskip10pt The next thing is to convert the values of the counters
into grayscale tones. Here it is convenient to make grayscale
proportional to $\log(c[i][j])$ rather than directly to $c[i][j],$
so as to be able to discern more details. In this case it is
necessary to initialize the counters to the starting value of $1,$
rather than to $0.$ That is how Fig.\ \ref{fig:Cube} was
created.\footnote{It is advisable to skip first $100--1000$
points, so that the point $\bx$ sets well on the attractor set,
but in practice the difference is undetectable with the eye.}
Fig.\ \ref{fig:octa}, was created using a similar method, for six
vertices of the regular octahedron, and using
$\alpha=0.4,0.5,0.6,0.7,0.8$ and $0.9,$ but with the help of
CLUCalc Visual Calculator, developed by Christian B.U. Perwass
\cite{clucalc}.
\section{From Quantum Fractals to Clifford algebras and beyond}
 There are several deficiences of the standard quantum theory. For instance:
\begin{enumerate}
  \item Need for external interpretation of the formalism
  \item Need for an ``observation"
  \item Two kinds of evolution: deterministic one, formalized by
  the Schr\"o\-din\-ger equation and ``projection postulate" of not so
  clear status (what constitutes a measurement?)
  \item Dubious role of time in Quantum Mechanics
  \item Paradoxes, like that of Schr\"odinger cat
  \item Impossibility of computer simulation of Reality (wave packet
  motion is not the only reality we want to explain)
\end{enumerate}
It is striking that the concept of an ``event" - which was of
crucial importance in creating special and general theories of
relativity finds no place in quantum formalism:

\begin{enumerate}
\item Barut \cite{barut90}, Bell \cite{bell89,bell90}, Chew \cite{chew85},
Haag \cite{haag90,haag92}, Shimony \cite{shimony86}, Stapp
\cite{stapp75,stapp77,stapp83,stapp93} and others stressed the
inadequacy of the Standard Quantum Theory for describing real-time
events
\item New technology enabled us to make continuous observations of
individual quantum systems. These experiments give us time series
of data - thus series of events and  not only the expectation
values (they may be ultimately computed)
\item What we observe are ``events". What we need to find and to
explain are regularities in time series of events.
\end{enumerate}
Einstein, Podolsky and Rosen \cite{einstein35} concluded that ``the
description of reality as given by a wave function is not complete."
John Stewart Bell, one of the most renowned theoretical physicists,
\cite{bell87} argued: ``Either the wave function, as given by the
Schr\"odinger equation, is not everything, or it is not right.(...)
If, with Schr\"odinger, we reject extra variables, then we must
allow that his equation is not always right. I do not know that he
contemplated this conclusion, but it seems to me inescapable." One
year before his untimely and premature death, Bell wrote these
insightful words in the paper that was his contribution to the
Conference ``62 Years of Uncertainty" held in Erice, Italy
\cite{bell90}:\begin{quotation}The first charge against
``measurement", in the fundamental axioms of quantum mechanics, is
that it anchors there the shifty split of the world into ``system"
and ``apparatus". A second charge is that the word comes loaded with
meaning from everyday life, meaning which is entirely inappropriate
in the quantum context. When it is said that something is
``measured" it is difficult not to think of the result as referring
to some preexisting property of the object in question. This is to
disregard Bohr's insistence that in quantum phenomena the apparatus
as well as the system is essentially involved. If it were not so,
how could we understand, for example, that ``measurement'' of a
component of ``angular momentum''\ldots in an arbitrarily chosen
direction\ldots yields one of a discrete set of values? When one
forgets the role of the apparatus, as the word ``measurement'' makes
all too likely, one despairs of ordinary logic.... hence ``quantum
logic". When one remembers the role of the apparatus, ordinary logic
is just fine.

In other contexts, physicists have been able to take words from
everyday language and use them as technical terms with no great harm
done. Take for example the ``strangeness", ``charm", and ``beauty"
of elementary particle physics. No one is taken in by this ``baby
talk".... as Bruno Touschek called it. Would that it were so with
``measurement". But in fact the word has had such a damaging effect
on the discussion, that I think it should now be banned altogether
in quantum mechanics.\end{quotation} Bogdan Mielnik
\cite{mielnik94}, analyzing the ``screen problem" - that is the
event of a quantum particle hitting the screen - noticed that ``The
statistical interpretation of the quantum mechanical wave packet
contains a gap", which he specified as ``The missing element of the
statistical interpretation: for a normalized wave packet $\psi({\bf
x} ,0)$ one ignores the probability of absorption on the surface of
the waiting screen. The time coordinate of the event of absorption
is not even statistically defined." John Archibald Wheeler
\cite{wheeler79} wrote: ``no elementary phenomenon is a phenomenon
until it is a recorded phenomenon." Eugene Wigner \cite{wigner86}
(see also \cite{freire1} for an overview of Wigner's position)
noticed that ``there may be a fundamental distinction between
microscopic and macroscopic systems, between the objects within
quantum mechanics' validity and the measuring objects that verify
the statements of the theory." Brian Josephson \cite{josephson01}
suggested that 'the observer' is a system that, while lying outside
the descriptive capacities of quantum mechanics, creates observable
phenomena such as wave function collapse through its probing
activities. Better understanding of such processes may pave the way
to new science."

Motivated by these and other similar conclusions of many authors I
decided to look for a ``way out of the quantum trap". While the real
solution may need a radical departure from the present scheme of
thinking about ``Reality", possible paths towards a better formalism
than the standard one have been investigated by many authors, mainly
along two lines. One is so called ``Bohmian mechanics", conceived
originally by Louis de Broglie as ``the theory of the double
solution" \cite{debroglie}, and then reformulated and developed by
David Bohm \cite{bohmhiley} (see also \cite{durr,hiley} for more
recent reviews, and \cite{freire2} for an interesting historical
overview). The other is known as the GRW (Ghirardi--Rimini--Weber)
or ``spontaneous localization model" (see \cite{GRW,bg1}). In
\cite{tom1,tom2} the GRW model has been generalized  so as to apply
not only to quantum mechanics, but also to  quantum field theory
(see also \cite{tom3} for a recent comparison between the two
approaches).

A further generalization of spontaneous localization theories has
been described in \cite{jad06a}, where a general formal structure of
quantum theories that incorporate the concept of events has been
formulated. This latter generalization enables us to define
precisely the very concepts of ``measurement" and ``experiment",
along the paths suggested by John Bell, and to model simultaneous
measurements of several non--commuting observables, in spite of the
warnings of standard quantum mechanical textbooks claiming that such
measurements contradict the very principles of quantum mechanics. As
this subject is directly related to the main topic of this paper
(the M\"obius transformations $\phi_\bp,\, \phi_\bq$ commute, only
if $\bp$ and $\bq$ are parallel or antiparallel) , some introduction
into the subject is given below.

The standard quantum theory, as formalized, for instance, by John
von Neumann \cite{neumann}, was based on postulates, and on
mathematical consequences derived from these postulates. The
postulates were to a large extent arbitrary, and other systems of
postulates have been proposed and discussed in the literature. Also
the physical interpretation of the mathematical results is not
unambiguous.

One of the most celebrated consequences of the quantum formalism is
the so called Heisenberg's uncertainty principle. Formally it states
that in any quantum state the product $\Delta_\psi^2(x)\times
\Delta_\psi^2(p_x)$ of the mean square deviation from the mean
values of the same components of the position and of the momentum
variables are bounded from below by $\hbar^2/4.$ This formal result
was, unfortunately, interpreted as an ``impossibility of a
simultaneous measurement of the position and momentum'', and, more
generally, of any pair of complementary, non--commuting observables.
I say ``unfortunately'', because while it is true that
non--commuting operators do not have, in  general, a joint
probability distribution, it has little to do with the possibility
or impossibility of performing their simultaneous measurements; the
main reason being that the concept of a ``measurement'' is not
defined within the formal framework of the standard quantum theory.

To define the measurement an extension or a revision of the quantum
theory is needed. The simplest extension is by using an algebraic
formulation but, at the same time, abandoning the standard
interpretation scheme. Let $\mathfrak{A}$ be an involutive algebra
over $\BR $ or $\BC,$ (for instance a $C^*$ or a von Neumann
algebra), and let $\mathfrak{Z}$ be its center. When $\mathfrak{Z}$
is trivial (that is when it consists of scalars only), then
$\mathfrak{A}$ is called a {\em factor\/} \cite[Chapter
V.1]{connes1}. A general algebra can be, essentially uniquely,
decomposed into a direct integral (or a direct sum) of factors
\cite[Theorem 8, p. 452]{connes1}:\begin{theorem} Let $\mathfrak{A}$
be a von Neumann algebra on a separable Hilbert space. Then
$\mathfrak{A}$ is algebraically isomorphic to a direct integral of
factors $$\int_X \mathfrak{A}(t)\, d\mu (t).$$\end{theorem}
\noindent Connes' comment on this decomposition theorem is worth
quoting:
\begin{quotation} ``This theorem of von Neumann shows that the factors
already contain what is original in all of the von Neumann algebras:
they suffice to reconstruct every von Neumann algebra as a
`generalized' direct sum of factors."\end{quotation} Although
formally correct, the statement above is, at the same time,
misleading. Every separable Hilbert space is a direct sum of
one--dimensional spaces. But that does not mean that
one--dimensional spaces contain what is original in all Hilbert
spaces. For instance non--commutativity shows up only when the
dimension of the Hilbert space is greater than one, and canonical
commutation relations, so important in physics, can be realized only
when the dimension of the Hilbert space is infinite; similarly with
algebras.

In quantum theory it is usually assumed that the relevant algebras
are factors. But, to include the ``events", to describe
``measurements", we need to go beyond that; we need to use more
general algebras, with a non--trivial center. This step allows us,
at the same time, to describe simultaneous ``measurement" of several
non--commuting observables. While there is no joint probability
distribution, the process is well defined and leads to chaos and to
fractal--like patterns, as seen, for instance, in Fig.
\ref{fig:octa} (see \cite{jad05a} for a comprehensive discussion of
this issue).

The crucial issue here is illustrated by the double role of the maps
$\phi_q$ (\ref{eq:1}). On on hand they are represented as belonging
to the group $Spin(1,4)$ and therefore they are (inner)
automorphisms of the Clifford algebra $C(E_{(1,4)}).$ On the other
hand they are represented as linear, positivity preserving,
transformations (see Eq. (\ref{eq:axa}))of the complex algebra
$\BC(2)$ of $2\times 2$ complex matrices. The maps $X\mapsto
X^\prime$ in Eq. (\ref{eq:axa}) are not automorphisms, therefore
they do not map central elements into central elements (even if the
center is trivial in this particular case), yet they preserve
positivity. It is positivity that is important in physical
applications, because it relates to the positivity of probabilities.

Quantum mechanics has been, originally, formulated as a theory over
the field of complex numbers. But there is no reason why it has to
be so. The fields of real numbers and of quaternions lead to
theories that are much like the standard quantum theory, except that
the domains of application of these alternatives are not yet known.

The statistical interpretation of the standard quantum mechanics is
based on the idea that the complex lines in a complex Hilbert space
describe ``pure states" of the system. But it does not have to be
so. Other schemes are possible; any positive cone can serve as a
statistical figure, and the probabilistic interpretation can result
from dynamics (like in the simple IFS system discussed in this paper
and in \cite{jad06a}, see also \cite{los} for a different approach
to ``Quantum Iterated Function Systems"), rather than be postulated.
This opens the way towards generalization of the quantum mechanical
framework and to a possible unification of quantum theory with
relativity, a unification that has been sought for more than 70
years. Clifford algebras, and closely related  CAR algebras
(Canonical Anticommutation Relations), and their generalizations,
provide one possible path. But there is also another path, going
beyond algebras based on binary operations. First steps in this
promising new direction have been taken by Frank D. Smith
\cite{smith93} and Yaakov Friedman \cite{yaakov05} (see also
\cite{satake} for the relevant mathematical background)

\textbf{Acknowledgements}: Thanks are due to the late Pertti
Lounesto for pointing to me the path relating quantum jumps and
conformal maps. I also thank Pierre Angl\`{e}s for discussions and
encouragement, and to QFG for a partial support of this work.
$\allowbreak$

{\small \vskip 1pc {\obeylines }}

{\small \noindent Arkadiusz Jadczyk }

{\small \noindent IMP }

{\small \noindent Chateau Saint Martin }

{\small \noindent Saint--Martin--Belcasse }

{\small \noindent 82100 Castelsarrasin}

{\small \noindent France }

{\small \vskip 1pc }

{\small \noindent E-mail: arkadiusz.jadczyk@cict.fr }

\end{document}